\newcommand{\be}{\begin{eqnarray}}
\newcommand{\ee}{\end{eqnarray}}
\documentstyle[preprint,aps]{revtex}
\begin{document}
\draft
\tightenlines
\title{Density and expansion effects on pion spectra in relativistic
       heavy-ion collisions}
\author{Alejandro Ayala and Julio Barreiro}
\address{Instituto de Ciencias Nucleares,\\
         Universidad Nacional Aut\'onoma de M\'exico,\\
         Aptdo.\ Postal 70-543, M\'exico D.F. 04510.}
\author{Luis M. Monta\~no}
\address{Centro de Investigaciones y de Estudios Avanzados,\\
         Aptdo. Postal 14-740, M\'exico D.F. 07000.}
\maketitle
\begin{abstract}

We compute the pion inclusive momentum distribution in heavy-ion collisions
at AGS energies, assuming thermal equilibrium and accounting for density
and expansion effects at the time of decoupling. We compare to data on
mid rapidity charged pions produced in central Au + Au collisions and find
a very good agreement. The shape of the distribution at low $m_t-m$ is 
explained in part as an effect arising from the high mean pion density achieved
in these reactions. The difference between the positive and
negative pion distributions in the same region is attributed in part to the 
different average yields of each kind of charged pions.

\end{abstract}

A great deal of experimental effort has been devoted in recent years to the 
production of a highly compressed state of matter in high-energy collisions 
between heavy ions. It has been speculated that at sufficiently high baryonic 
densities or temperatures, a phase transition from hadronic matter to a 
quark-gluon plasma can occur. The study of particle spectra should provide
useful information about the dynamics and evolution of the kind of matter 
formed in these reactions.

Some of the most copiously produced particles in central high-energy heavy ion 
collisions are pions. An understanding of pion production during these 
collisions has long been searched, particularly in view of one of the most 
remarkable properties exhibited by their spectra~\cite{exper}, commonly 
referred to as an enhancement of either the low or the high transverse momentum
regions in the inclusive single pion distribution, as compared to p-p 
collisions. This property is concomitant with the difficulty to describe the 
invariant transverse mass distribution with a single exponential 
function~\cite{E802}. 
 
Several possibilities have been put forward to explain the peculiar shape of 
the distributions, amongst them, the different contributions to the 
pion yield coming from the decay of $\Delta$ resonances produced during 
different stages of the collision~\cite{Li} and the superposition of primary 
pions and pions coming from resonance decay, mainly $\Delta$'s~\cite{Brown}.
The importance of transverse flow in describing the spectra has also been 
stressed~\cite{Atwater}. The proper treatment of Coulomb final state 
interactions has also been pointed out. More recently, Wong and 
Mostafa~\cite{Wong} noticed that when particles feel the effects of a boundary 
during the evolution from the first stages of the collision to their final 
free streaming, their momentum distribution is affected due to the 
discretization of energy levels and the correspondingly different density of 
states introduced by the finite size of the system just before freeze out. 
This idea is based on the concept of a pion liquid, first discussed by 
Shuryak~\cite{Shuryak} and was further developed in Ref.~\cite{Ayala} with the 
introduction of a finite chemical potential associated to the mean pion 
multiplicity per event in central collisions. 

Calculations based on the effects of a boundary resort to the assumption of 
thermal equilibration. They are successful in reproducing the concave shape of 
the distribution at high transverse momentum but fail to describe its overall 
fall off. This failure could have been anticipated since, as pointed out
in Ref.~\cite{Landau}, thermodynamics is tantamount of hydrodynamics and
a simple thermal spectrum should be corrected by the Doppler shift resulting 
from collective expansion. A challenge remained as to how to incorporate 
hydrodynamical effects into a description of the transverse momentum 
distribution based on discrete energy states. In this work we meet this 
challenge and lay down the formalism required to account simultaneously for
expansion and boundary effects in a phenomenological calculation of pion 
spectra in relativistic heavy-ion collisions. We apply this formalism to 
compute the invariant transverse momentum distribution for mid rapidity pions 
at AGS energies. We find a very good agreement with data on central Au+Au 
reactions at $11.6~A$~GeV/c. Further distortions of the spectra arising from
Coulomb effects will be discussed in an upcoming work.

In order to collect the necessary ingredients for the calculation, recall that 
if the average pion separation $d$ ever becomes smaller than the average
range of the pion strong interaction $d_s$ ($\sim 1.4$~fm) during the 
evolution of the collision, the pion dispersion curve can be modified and the 
collective properties of the pion system resemble those of a liquid rather than
those of a gas~\cite{Shuryak}. An important consequence is the appearance of a 
surface tension that acts as a reflecting boundary. The E-802/866 collaboration
has reported that a baryon density of about eight times normal nuclear density 
is achieved in central Au+Au collisions at $11.6~A$~GeV/c. A large fraction of 
this density is due to pions. From Fig.~3 of Ref.~\cite{E802}, one can read 
that the total number of charged pions one unit around central rapidity in
this kind of reactions is about 200. Under the assumption that the number of
neutral pions in the same rapidity interval is half the total number of charged
ones, the total pion yield in central collisions at mid rapidity is 
$dN_{\pi}/dy\sim$~300. Since the average pion separation is inversely 
proportional to one third of the pion density $n_{\pi}=(1/At_0)dN_{\pi}/dy$, 
where $A$($\sim 64$~fm$^2$) is the transverse area of the reaction and 
$t_0$($\sim 1$~fm) is a typical formation time~\cite{Bjorken}, 
$d\sim 0.6$~fm $<d_s$ and thus the condition to regard the pion system as 
a liquid is met. 

Pions that move towards the boundary of the system feel the attractive
potential behind them and are reflected back. The reflection details depend on 
the wave length of the given particle but the property introduced by the 
reflecting surface is that it allows very little wave function {\it leakage}
and to a good approximation, the pion wave functions vanish outside
the boundary. When, as a consequence of the expansion of the initially 
compressed and hot system, the pion average separation becomes larger than 
the range of strong interactions, the system becomes a free gas but the 
transition between the liquid and the gas phases is very rapid and the pion 
momentum distribution should be determined by the distribution just before 
freeze out.

Thus the system of pions can be considered as confined and their wave
functions as satisfying a given condition at the boundary just before freeze 
out. In this case, the energy states form a discrete set. The shape of
the volume within the confining boundary deserves some attention. For reactions
with a large degree of transparency, Bjorken like geometry, with a 
predominantly longitudinal elongation, seems better suited. However, for AGS 
energies, a significant amount of stopping has been reported by the E-802/866 
collaboration in central Au+Au reactions\cite{E802}. In this case, a more 
symmetric geometry between transverse and longitudinal directions seems 
appropriate. We thus consider a scenario in which the system of pions of
a given species is in thermal equilibrium and is confined within a sphere
of radius $R$ (fireball) as viewed from the center of mass of the colliding
nuclei at the time of decoupling. As discussed in Ref.~\cite{Kusnezov}, this
time needs not be the same over the entire reaction volume. Nevertheless, 
in the spirit of the fireball model of Ref.~\cite{Siemens}, we consider that 
decoupling takes place over a constant time surface in space-time. This 
assumption should be essentially correct if the freeze out interval is short 
compared to the system's life time. 

In the absence of expansion, the solution has been found in Ref.~\cite{Ayala}.
This involves solving the Klein-Gordon equation to find the stationary
wave functions satisfying 
\be
   \left[ \frac{\partial^2}{\partial t^2} - \nabla^2 + m^2\right]
   \psi({\mathbf r},t)=0\, ,
   \label{eq:oldsol}
\ee
subject to the condition
\be
   \psi(|{\mathbf r}|=R,t)=0\, ,
   \label{eq:cond}
\ee
and finite at the origin. In order to incorporate the effects of a 
hydrodynamical flow, we observe that the presence of an ordered motion, 
represented by a four-velocity field $u^{\mu}=\gamma(r)(1,{\mathbf v}(r))$, 
amounts to a redistribution of momentum in each of the fluid cells, as viewed 
from a given reference frame (the center of mass in our case). The tendency of 
matter to occupy a larger volume is compensated by the distribution of momenta 
in each cell becoming narrower\cite{Siemens}. The distribution in the cell 
becomes also centered around the momentum associated with the velocity of the 
fluid element. Consequently, the thermal spectrum in each cell should be 
described on top of this collective flow, that is, referred from the collective
fluid's element momentum. To describe this behavior of the expanding, bound 
system of pions, we make the substitution of the momentum operator $p^{\mu}$ 
by $p^{\mu} - m u^{\mu}$, where $m$ is the pion mass. The term $m u^{\mu}$ 
represents the collective momentum of the given pion fluid element. 
The corresponding equation becomes
\be
   \left[ -\Big( i\frac{\partial}{\partial t}-m\gamma(r) \Big)^2+\Big(
   -i{\mathbf \nabla} -m\gamma(r){\mathbf v}(r) \Big)^2 + m^2\right]
   \psi({\mathbf r},t)=0
   \label{eq:neweq}
\ee
and we look for stationary solutions subject to the same condition in 
Eq.~(\ref{eq:cond}) and also finite at the origin. We consider a 
parametrization of the three velocity vector ${\mathbf v}(r)$ that scales
with the distance from the center of the fireball.
\be
   {\mathbf v}(r)=\beta\frac{r}{R}\hat{\mathbf r}\, ,
   \label{eq:vel}
\ee
We identify this velocity with the transverse flow velocity and ignore any 
asymmetry between transverse and longitudinal expansion. $0<\beta <1$, 
represents the surface fireball velocity. The corresponding explicit 
expression for $\gamma (r)$ is
\be
   \gamma (r)=\frac{1}{\sqrt{1-\beta^2 r^2/R^2}}\, .
   \label{eq:gam}
\ee 
Eq.~(\ref{eq:neweq}) with the gamma factor given by Eq.~(\ref{eq:gam}) can
only be solved numerically. In order to provide an analytical solution,
we approximate the function $\gamma$ by the first terms of its Taylor 
expansion
\be
   \gamma (r)\simeq 1+\frac{\beta^2}{2}\frac{r^2}{R^2}\, .
   \label{eq:gamap}
\ee
This approximation is valid for not too large values of $\beta$. In this
case, Eq.~(\ref{eq:neweq}) becomes an equation for particles moving in a 
spherical harmonic well with a rigid boundary.

The stationary states are
\be
   \psi_{nlm'}({\mathbf r},t)&=&\frac{A_{nl}}{\sqrt{2E_{nl}}}e^{-iE_{nl}t}
   e^{im\beta r^2/(2R)}Y_{lm'}(\hat{\mathbf r})\nonumber \\
   &\times& e^{-\alpha_{nl}^2r^2/2}
   r^l\,_1F_1\left(\frac{(l+3/2)}{2}-\frac{\varepsilon_{nl}^2}{4\alpha_{nl}^2},
   l+3/2;\alpha_{nl}^2r^2\right)\, ,
   \label{eq:solnew}
\ee
where $\,_1F_1$ is a confluent hypergeometric function and $Y_{lm'}$ is a 
spherical harmonic. The quantities $A_{nl}$ are the normalization constants 
and are found from the condition
\be
   \int d^3r \psi^{\ast}_{nlm'}({\mathbf r},t)
   \frac{\stackrel{\leftrightarrow}{\partial}}{\partial t}
   \psi_{nlm'}({\mathbf r},t) = 1\, .
   \label{eq:norma}
\ee
The parameters $\alpha_{nl}$ and $\varepsilon_{nl}$ are related to the energy 
eigenvalues $E_{nl}$ by
\be
   \alpha^4_{nl}&=&m(E_{nl}-m)\beta^2/R^2\, ,\nonumber\\
   \varepsilon^2_{nl}&=&E_{nl}(E_{nl}-2m)\, .
   \label{eq:param}
\ee
$E_{nl}$ are given as the solutions to
\be
   \,_1F_1\left(\frac{(l+3/2)}{2}-\frac{\varepsilon_{nl}^2}{4\alpha^2_{nl}},
   l+3/2;\alpha_{nl}^2R^2\right)=0\, .
   \label{eq:encond}
\ee
The normalized contribution to the momentum distribution from the energy
state with quantum numbers $n,l,m'$ is given in terms of the absolute value
squared of the Fourier transform of Eq.~(\ref{eq:solnew}), namely,
\be
   \psi_{nlm'}({\mathbf p})=\int d^3r e^{-i{\mathbf p}\cdot{\mathbf r}}
   \psi_{nlm'}({\mathbf r})\, .
   \label{eq:tnas}
\ee
Since the problem has an azimuthal symmetry, the wave function in momentum
space does not depend on the quantum number $m'$ and is a function only of the
momentum magnitude.
\be
   \psi_{nlm'}({\mathbf p})=\psi_{nlm'}(p)\delta_{m'\,0}\, .
   \label{eq:onlymag}
\ee
Consequently, the momentum distribution is obtained by weighing the 
contribution from each state with the statistical Bose--Einstein factor and
adding up the contribution from all of the states
\be
   \frac{d^3N}{d^3p}=\sum_{n,\,l}\frac{\phi_{nl}(p)}{e^{(E_{nl}-\mu)/T}-1}
   \label{eq:dist}
\ee
where $\phi_{nl}(p)$ is defined by
\be
   \phi_{nl}(p)=\frac{2E_{nl}}{(2\pi)^3}|\delta_{m'0}\psi_{nlm'}(p)|^2
   \label{eq:nwedefphi}
\ee
and the chemical potential $\mu$ is computed from
\be
   N=\sum_{n,\,l}\frac{(2l+1)}{e^{(E_{nl}-\mu)/T}-1}\, ,
   \label{eq:chem}
\ee
for a given number of particles $N$. Eq.~(\ref{eq:chem}) follows from
Eq.~(\ref{eq:dist}) after integration over $d^3p$.

Fig.~1 shows the systematics obtained by varying the parameters involved.
The curves in Figs.~1$a$, $b$ and $c$ are computed for mid rapidity pions, 
$y=0$, for which the assumption of spherical expansion should not be important 
since these do not experience the effect of longitudinal flow. Fig.~1a shows 
the behavior of the distribution for a fixed temperature $T=120$ MeV, a fixed 
value of the surface expansion velocity $\beta =0.5$ and a total number of 
particles $N=150$ for various values of the fireball's radius $R$. Notice the 
convex shape of the distribution at low $m_t-m$ for large values of $R$ and the 
transition to a concave shape with decreasing values of $R$. This is a density 
effect since for large $R$ and a fixed total number of particles the density 
is lower than for smaller values of $R$. In the former case, the value taken by
the chemical potential is far from the first energy state, whereas in the 
latter, the chemical potential is close to the energy of this state and thus 
the lowest lying energy states contribute with a more significant statistical 
weight. The same effect can be obtained by keeping a fixed radius and varying
the number of particles.

Fig.~1b shows the behavior of the distribution when varying the temperature $T$
maintaining fixed values of the radius $R=8$ fm, the surface expansion 
velocity $\beta =0.5$ and the total number of particles $N=150$. As could be 
expected, the main effect goes into the effective slopes describing
the distribution's overall fall off. The distribution for $T=100$ MeV
rises more steeply at low values of $m_t-m$, due to the
proximity of the system to the critical temperature for Bose-Einstein
condensation, as discussed in Ref.~\cite{Ayala}. Fig.~1c shows the 
effect of varying the surface expansion velocity comparing the cases with 
$\beta =0$ and $\beta =0.5$ keeping fixed the values of the radius $R=8$ fm, 
the temperature $T=120$ MeV and the total number of particles $N=150$. Notice 
that the effect corresponds also to assigning different overall effective 
inverse slopes to each curve, the largest one corresponding to the curve 
with non-vanishing expansion velocity. Finally, Fig.~1d shows the shape of
the distribution for a rapidity away from the central region, in this
case with the parameters $T=120$ MeV, $\beta =0.5$, $R=8$ fm and $N=150$ for 
$y_{lab}=3.0$. Notice the pronounced bending upwards of the distribution
at large values of $m_t-m$. This effect can be attributed to a larger
density of states at high energy eigenvalues as compared to a calculation 
without boundary.

With these systematics at hand, we proceed to describe the mid rapidity
pion data on central Au+Au reactions at $11.6~A$~GeV/c.\cite{E802}. 
Data correspond to pions within a rapidity interval $|\Delta y|< 1$
around central rapidity. Given our assumption of a spherically symmetric
fireball, our calculation will compare best to this part of the spectrum
since these are the pions that do not experience the effects of a longitudinal
flow that might be different from the flow in the transverse direction.

Rather than performing an exhaustive search in the whole parameter space and
in order to test the plausibility of this type of description, here we fix the 
value of the parameters involved to reasonable and more or less accepted 
values. It is clear that a more complete analysis requires the proper
treatment of Coulomb effects, this will be the subject of an upcoming work.
We take~\cite{Esumi} $T=120$ MeV, $\beta =0.5$ (corresponding to an average 
collective expansion velocity $<v>\simeq 0.4$c). For the mean negative pion
multiplicity we take $N_{\pi^-}=160$~\cite{Videbaek}. We consider a fireball 
radius $R=8$ fm. Fig.~2a shows the theoretical distribution compared to 
data. In order to compare with the invariant differential cross section  
reported in Ref.\cite{E802} which is normalized to a subset of (central
rapidity) 116 negative pions, the curve has been multiplied by a constant 
${\mathcal N}=0.56$ that minimizes the $\chi^2$ when we compare to data above 
$m_t-m=0.4$ GeV, the region where Coulomb effects should start becoming less 
significant. Above $m_t-m=0.4$ GeV the agreement between data and theory is 
very good. Below $m_t-m=0.4$ GeV the curve follows the shape of the data points
but these last are still above the calculation. This could be a good feature 
since one knows that the long-range Coulomb effects should push the 
distribution for low momentum negative pions upwards, given that their Coulomb 
interaction with the overall positive charge is attractive.

We now use these parameters to describe data on positive pions. Fig.~2b shows 
the theoretical distribution calculated for $T=120$ MeV, $\beta =0.5$, $R=8$ fm
but a total positive pion multiplicity $N_{\pi^+}=115$~\cite{Videbaek}, 
compared to data. In order to compare with the invariant differential cross 
section reported in Ref.~\cite{E802} which is normalized to a subset of 
(central rapidity) 94 positive pions, the curve has been multiplied by the 
constant ${\mathcal N}=0.59$ that minimizes the $\chi^2$ when we compare to 
data above $m_t-m=0.4$ GeV. The agreement between data and theory is also very 
good for the region above $m_t-m=0.4$ GeV. However, the raise of the curve 
below $m_t-m=0.4$ GeV is less steep than for the negative pion case. This is 
easy to understand since for positive pions the density is lower than for the 
negative ones. Also in the same region, the theoretical curve is marginally
below data and the Coulomb distortion will push it even more below. We
speculate that this signals that data prefer a slightly lower value of the 
radius but again, this can only be confirmed after the proper inclusion of 
Coulomb effects.

In conclusion, we have shown that a very good description of mid rapidity,
charged pion spectra can be achieved by a phenomenological calculation whose
key ingredient is the proper treatment of the large pion density produced in 
central, relativistic Au+Au reactions. This large density leads to consider
the pion system as confined during the early stages of the collision, before
decoupling. Such scheme has consequences on both ends of the spectra. At low 
transverse mass, the convex shape of the distribution is due to the large 
value of the chemical potential associated with the mean pion multiplicity 
per event. At high transverse momentum, the convex shape of the distribution 
is due to the higher density of states as compared to a calculation without 
boundary. Another important element is the inclusion of collective flow. We 
find that an average flow velocity of $<v>\simeq 0.4$c together with a 
temperature $T=120$ MeV does a very good job describing data above 
$m_t-m=0.4$ GeV when the fireball's radius is about $R=8$ fm. A more conclusive
statement can be made only after we include a proper treatment of the Coulomb 
corrections~\cite{Ayala2}. Perhaps more importantly is the fact that the 
different rates at which the positive and negative pion distributions rise at 
low values of $m_t-m$ can be understood in part as an effect related to
the correspondingly different measured yields, making the negative pion 
subsystem denser that the positive pion one. 
  
Support for this work has been received in part by CONACyT M\'exico under grant
No. I27212-E.

\section*{Figure Captions}

\noindent
Fig.~1: Systematics obtained by varying $\mathsf{(a)}$ the fireball's 
radius $R$, $\mathsf{(b)}$ the temperature $T$ and $\mathsf{(c)}$ the surface
expansion velocity $\beta$ for rapidity $y_{cm}=0$. $\mathsf{(d)}$ Distribution
for a rapidity $y_{lab}=3.0$. 

\noindent
Fig.~2: $\mathsf{(a)}$ Theoretical distribution $(2{\pi}m_t)^{-1}d^2N/dm_tdy$
computed  with the parameters $T=120$ MeV, $\beta=0.5$, $N_{\pi^-}=160$ and
$y_{cm}=0$ compared to data from the E-802/866 on mid rapidity negative pions
from central Au+Au reactions at $11.6 A$ GeV/c. The theoretical curve has been
multiplied by the constant ${\mathcal N}=0.56$ that minimizes the $\chi^2$ when
compared to data above $m_t-m=0.4$ GeV. $\mathsf{(b)}$ Distribution computed
with the same parameters but with $N_{\pi^+}=115$ compared to data on
mid rapidity positive pions from the same reaction. The  theoretical curve has
been multiplied by the constant ${\mathcal N}=0.59$ that  minimizes the
$\chi^2$ when compared to data above $m_t-m=0.4$ GeV. Data are
$(2{\pi}m_t\sigma_{trig})^{-1}d^2\sigma/dm_tdy$ in the rapidity interval
$0<\Delta y < 0.2 $. The total measured yield spans the rapidity
interval $| \Delta y | < 1$ around central rapidity~\cite{E802}. 


\begin{thebibliography} {20}

\bibitem{exper}
H. Str\"obele et al (NA35 Collaboration), Z. Phys. C {\bf 38} (1988) 89;
R. Albrecht et al (WA80 Collaboration), Z. Phys. C {\bf 47} (1990) 367;
A {\bf 590} (1995) 259c;
J. Barrette et al (E814 Collaboration), Phys, Lett. B {\bf 351} (1995) 93.

\bibitem{E802}
L. Ahle et al (E-802/866 Collaboration), Phys. Rev. C {\bf 57} (1998) R466. 

\bibitem{Li}
B.-A. Li and W. Bauer, Phys. Lett. B {\bf 254} (1991) 335.

\bibitem{Brown}
R. Brocmann et al, Phys. Rev. Lett. {\bf 53} (1984) 2012; 
J. Sollfrank, P. Koch and U. Heinz, Phys. Lett. B {\bf 252} (1990) 256;
G.E. Brown, J. Stachel and G.M. Welke, Phys. Lett. B {\bf 253} (1991) 19.

\bibitem{Atwater}
See for example T.W. Atwater, P.S. Freier and J.I. Kapusta, Phys. Lett. B 
{\bf 199} (1987) 30;
K.S. Lee and U. Heinz, Z. Phys. C {\bf 43} (1989) 425.

\bibitem{Wong}
C.-Y. Wong, Phys. Rev. C {\bf 48} (1993) 902;
M.G.-H. Mostafa and C.-Y. Wong, Phys. Rev. C {\bf 51} (1995) 2135.

\bibitem{Shuryak}
E.V. Shuryak, Phys. Rev. D {\bf 42} (1990) 1764.

\bibitem{Ayala}
A. Ayala and A. Smerzi, Phys. Lett. B {\bf 405} (1997) 20.

\bibitem{Landau}
L.D. Landau, Izv. Akad. Nauk SSSR Ser, Fiz. {\bf 17} (1953) 51 also in
{\it Collected papers of L.D. Landau} (Pergamon Press and Gordon and Breach, 
New York, 1965), pp 569;
E.V. Shuryak and O.V. Zhirov, Phys. Lett. {\bf 89B} (1980) 253.

\bibitem{Bjorken}
J.D. Bjorken, Phys. Rev. D {\bf 27} (1983) 140.

\bibitem{Kusnezov}
D. Kusnezov and G. Bertsch, Phys. Rev. C {\bf 40} (1989) 2075.

\bibitem{Siemens}
P.J. Siemens and J.O. Rasmussen, Phys. Rev. Lett. {\bf 42} (1979) 880.

\bibitem{Esumi} As discussed by S. Esumi, S. Chapman, H. van Hecke and
N. Xu, Phys. Rev. C {\bf 55} (1997) R2163, there exists an anticorrelation 
between the transverse flow velocity and the freeze out temperature in
such a way that higher temperatures imply lower expansion velocities and
visce versa. It is likely that for AGS energies not too high temperatures 
are reached and thus the combination of $T$ and $\beta$ that we consider.

\bibitem{Videbaek}
F. Videbaek et al (E-802/866 Collaboration), 
Proc. Quark Matter 95, Nucl. Phys. A {\bf 590} (1995) 249c.
The E-802/866 Collaboration has reported a total number of negatively and
positively charged pions on the order of $N_{\pi^-}\simeq 160$ and 
$N_{\pi^+}\simeq 115$.

\bibitem{Ayala2}
A method to incorporate Coulomb corrections in the description of charged
pion spectra, that can be applicable in the context of this work, has
recently been proposed, see A. Ayala and J. Kapusta, Phys. Rev. C {\bf 56}
(1997) 407; A. Ayala, S. Jeon and J. Kapusta, nucl-th/9807062.

\end{thebibliography}
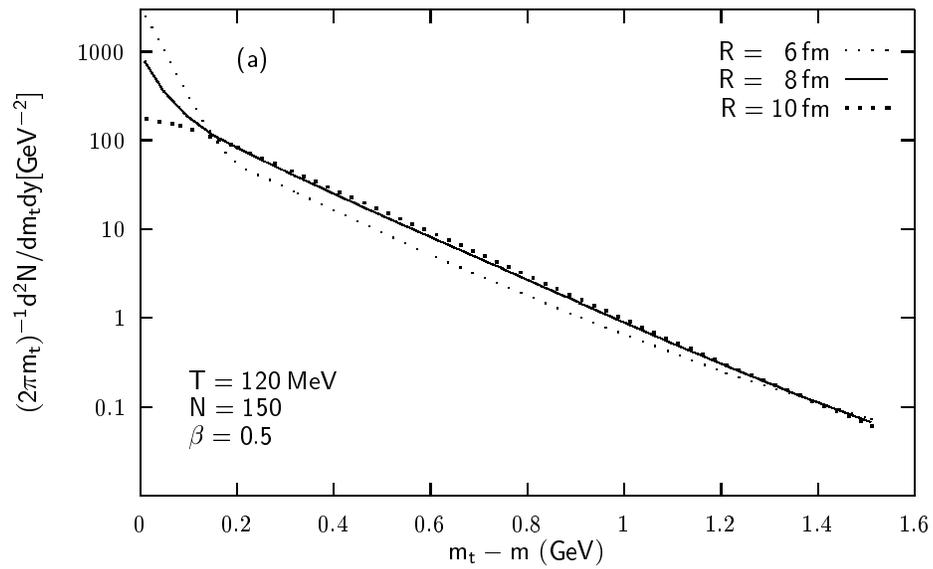
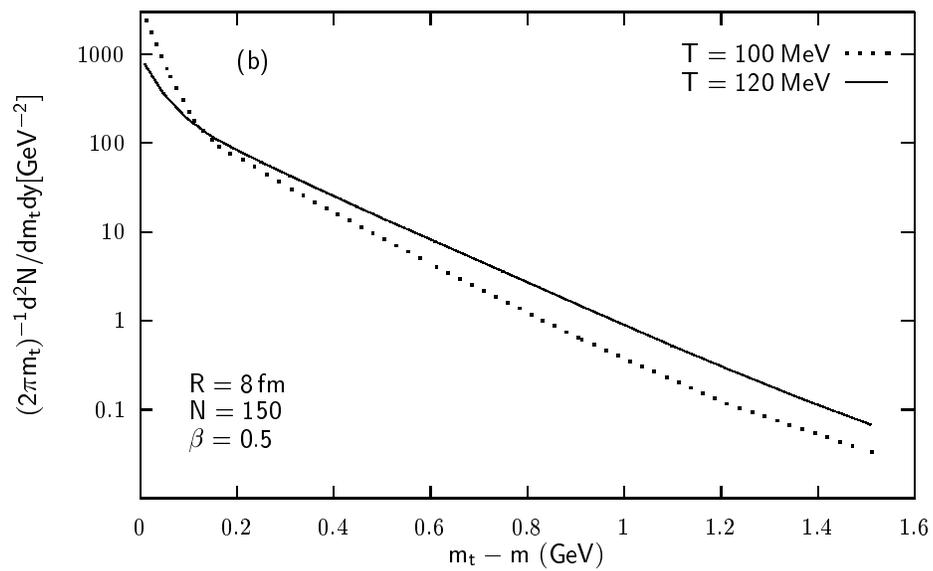
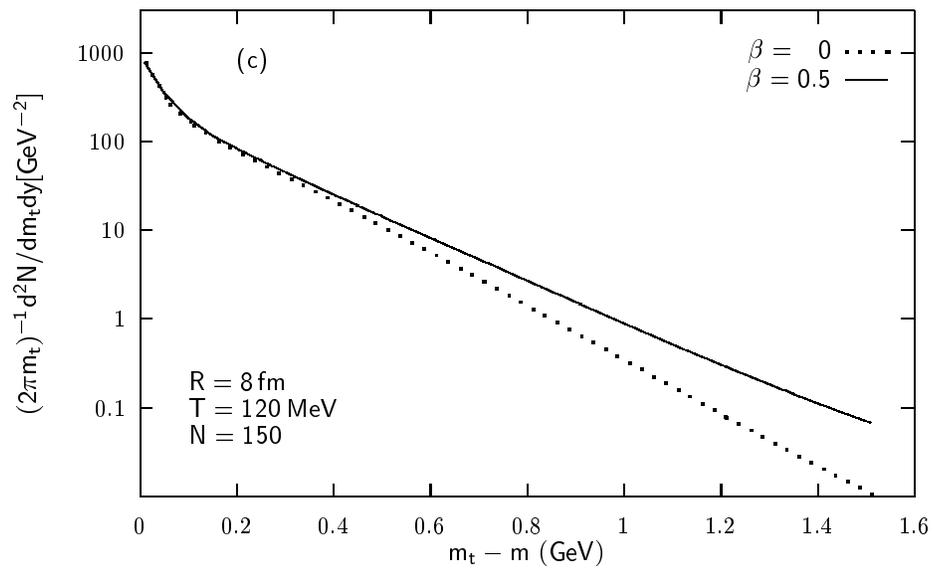
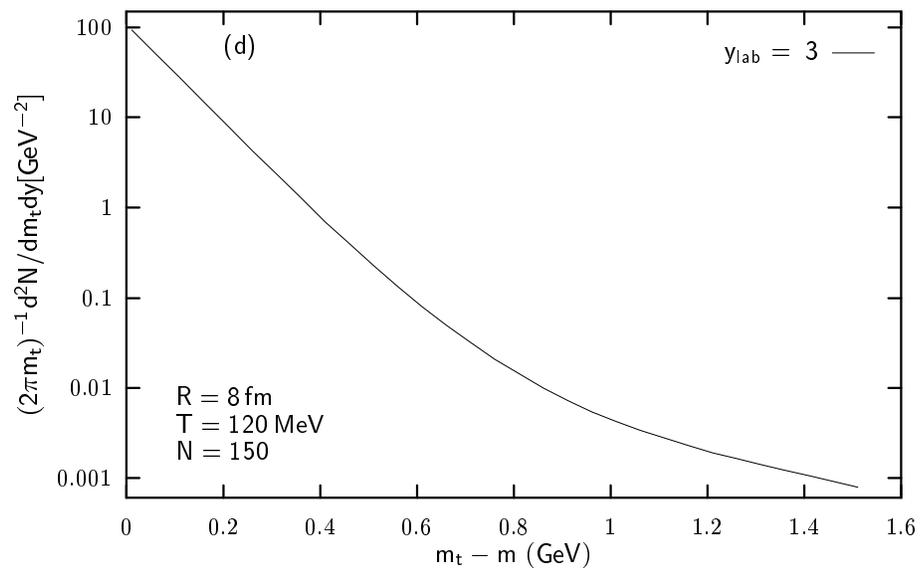
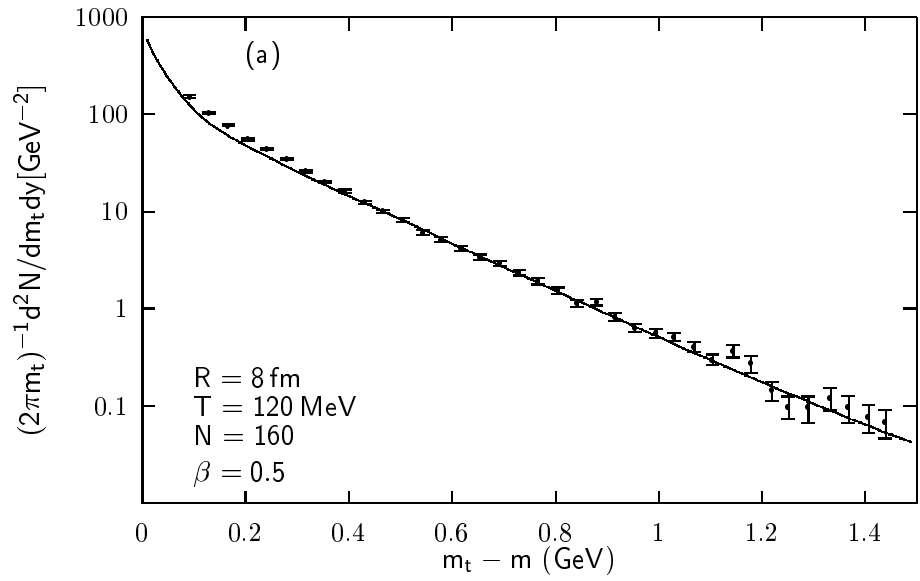
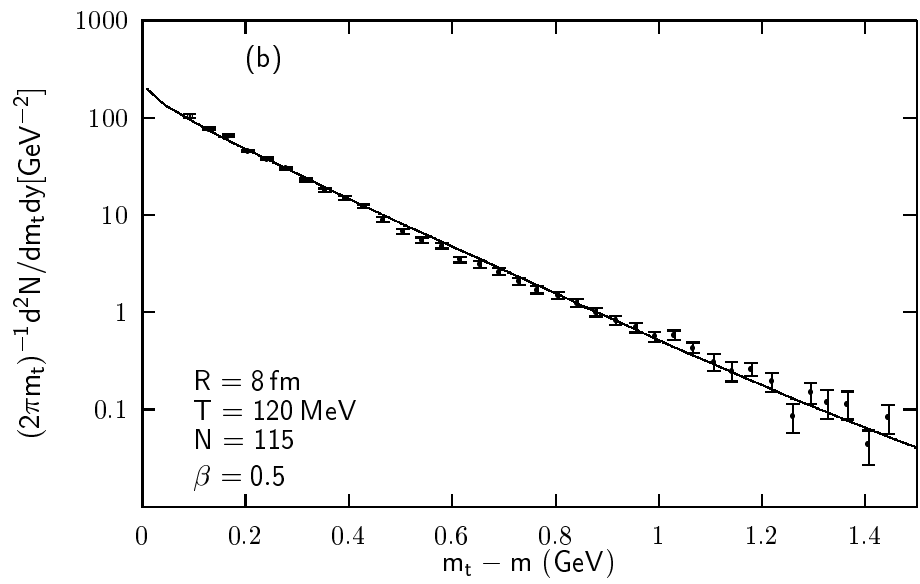
\end{document}